\documentclass[preprint,showpacs,preprintnumbers,amsmath,amssymb,nofootinbib]{revtex4}

\usepackage{graphicx}% Include figure files
\usepackage{dcolumn}% Align table columns on decimal point
\usepackage{bm}% bold math

\newcommand{\bi}{\bibitem}
\newcommand{\nn}{\nonumber}

\newcommand{\be}{\begin{eqnarray}}
\newcommand{\ee}{\end{eqnarray}}
\catcode`\@=11
\def\lsim{\mathrel{\mathpalette\@versim<}}
\def\gsim{\mathrel{\mathpalette\@versim>}}
\def\@versim#1#2{\vcenter{\offinterlineskip
\ialign{$\m@th#1\hfil##\hfil$\crcr#2\crcr\sim\crcr } }}
\catcode`\@=12

\begin{document}
\begin{center}
{\em To my late father}
\end{center}
\vspace{2cm}
\preprint{KANAZAWA-05-07}

\title{Dihedral Flavor Symmetry \\from Dimensional Deconstruction}

\author{Jisuke Kubo}

\affiliation{
Institute for Theoretical Physics, Kanazawa
University, Kanazawa 920-1192, Japan
\vspace{3cm}
}

\begin{abstract}
Extra dimension deconstructed
on a closed chain has  naturally the symmetry of a regular polygon,
the dihedral symmetry $D_N$.
We assume that the fields are irreducible representations of 
the binary dihedral group $Q_{2N}$,
which is the covering group of $D_N$.
It is found that although the orbifold boundary conditions
break the dihedral invariance explicitly,
the $Q_{2N}$ symmetry appears as an intact internal,
global flavor symmetry at low energies.
A concrete predictive model 
based on $Q_{6N}$ with an odd $N$ is given.

\vspace{3cm} 
\end{abstract}

\pacs{11.25.Mj,11.30.Hv, 12.15.Ff, 14.60.Pq, 02.20.Df }

\maketitle

\section{Introduction}
The Yukawa sector of the standard model (SM)
contains a large  number of  redundant parameters.
The presence of the  redundant parameters 
is not related to a symmetry in the SM. 
That is, they will appear in higher orders in perturbation theory
even if they are set equal to zero at the level.
These redundant parameters may become physical
parameters when going beyond the SM, and, moreover, they can induce
flavor changing neutral currents (FCNCs) and CP
violating phenomena that are absent or strongly suppressed
in the SM. One of the most well known examples is the
case of the minimal supersymmetric model (MSSM).
Since the SM can not control the redundant parameters,
the size of the new FCNCs and CP violating phases
 may be unacceptably  large unless
 there is some symmetry, or  one fine tunes  their values
 \footnote{For  recent reviews, see, for instance,
\cite{susy} and references therein. }.

A natural guidance to constrain the Yukawa sector
and to reduce the redundancy of this sector is a flavor symmetry.
It has been recently realized that 
nonabelian discrete flavor
symmetries, especially dihedral symmetries, can
not only reduce the redundancy,
but also partly explain the large mixing of neutrinos \footnote{
Models based on 
dihedral flavor symmetries,
ranging from $D3 (\simeq S_3)$ to $Q_6$ and $D_7$,
have been  recently discussed  in
 \cite{frampton1}-\cite{grimus4}.}.
When
supersymmetrized, it has been found that the same flavor symmetries can suppress FCNCs that are caused by soft supersymmetry
breaking terms \cite{kobayashi2,choi}(see also
\cite{hall2}-\cite{king2}).

In this letter we address the question of the origin of 
dihedral flavor symmetries.
We will find that dimensional 
deconstruction \cite{hamed1,hill} is a possible  origin of 
dihedral flavor symmetries.

\section{Dihedral invariance in an Extra 
dimensional space}
Consider an extra dimension
which is compactified on
a closed one-dimensional lattice with $N$ sites.
 We assume that the lattice has
the form of  a regular polygon with $N$ edges
as it is illustrated in Fig.~1.
\begin{center}
\begin{figure}[htb]
\includegraphics*[width=0.3\textwidth]{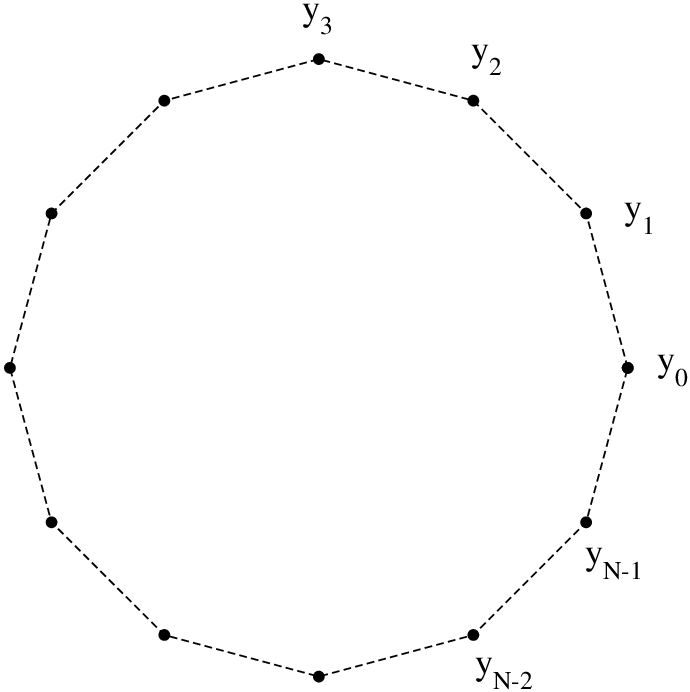}
\caption{\label{fig1}\footnotesize
A  regular polygon with $N=12$ edges, which are located at
$y=y_0,y_1,\dots,y_{N-1}$.}
\end{figure}
\end{center}
\normalsize
The  regular polygon is invariant under the symmetry
operations of the dihedral group $D_N$.
The $D_N$ operations are $2N$ discrete rotations, where $N$ of 
$2N$ rotations are combined with a parity transformation.
Clearly, a discrete polygon rotation of
$n\times \theta_N,~n \in\{1,\dots, N\}$
 corresponds to a discrete translation of the lattice sites
 of $n \times a$, where $a$ is the lattice spacing and
\be
 \theta_N &\equiv 2 \pi/N.
 \label{thetaN}
 \ee
The coordinate of the extra dimension is denoted by
$y$, and the $N$ sites are located at
$y=y_0,y_1,\dots,y_{N-1}$. ($y_{N+i}$ is identified with $y_{i}$.)
Under a $D_N$ transformation, 
the set of coordinates  $(y_0,y_1,\dots,y_{N-1})$
changes to  $(y_0',y_1',\dots,y_{N-1}')$, which we express in terms
of a $N\times N$ real matrix.
The matrix for the fundamental rotation (i.e.,
a rotation of $\theta_N$) is given by
\be
R_N &=&
  \left( \begin{array}{ccccc}
0  & 0 & 0 & \cdots &1 \\
1 & 0 & 0 & \cdots & 0\\
0 &  1  &0& \cdots & 0\\
 &   &\cdots &  & \\
0 &  0  &\cdots  & 1& 0\\
\end{array}\right),
\label{RN}
\ee
and that for the parity transformation is
\be
P_D  &=&
  \left( \begin{array}{ccccc}
1  & 0 &   \cdots & 0 &0 \\
0 & \cdots  & 0 &0 & 1\\
0 &   \cdots & 0&1&  0\\
 &   &\cdots &  & \\
0 &  1  & 0& \cdots & 0\\
\end{array}\right).
\label{PD}
\ee
Then the $2N$ group elements of $D_N$ are:
\be
{\cal G}_{D_N}=
\{R_N, (R_N)^2,\dots,(R_N)^N={\bf 1},
R_N P_D ,(R_N)^2 P_D, 
\dots, (R_N)^N P_D=P_D\}.
\ee
Using the properties,
$P_D^2={\bf 1}$ and $P_D R_N P_D=(R_N)^{-1}$,
one can convince oneself that ${\cal G}_{D_N}
$ is indeed  a group.

There exist two-dimensional representations 
for $\tilde{R}_N$ and $\tilde{P}_D$ \cite{frampton1,babu4}:
\begin{eqnarray}
\tilde{R}_N &=&
\left(\begin{array}{cc}
\cos \theta_N & \sin\theta_N  \\
-\sin \theta_N & \cos\theta_N  \end{array}\right)
\label{RN2},~
\tilde{P}_D = \left(\begin{array}{cc}1 & 0 \\ 0 & -1 \end{array}\right),
\label{PD2}
\end{eqnarray}
which are useful representations in finding irreducible representations
(irreps) of $D_N$ ($\theta_N$ is given in (\ref{thetaN})).
It follows that $D_N$ is a subgroup of $SO(3)$, which one sees
if one  embeds $\tilde{R}_N$ and $\tilde{P}_D$ into $3\times 3$ matrices
\be
\tilde{R}_N&\to & 
\left(\begin{array}{ccc}\cos \theta_N & 
\sin\theta_N &0 \\ -\sin \theta_N & \cos
\theta_N &0 \\ 0 & 0 &1 \end{array}\right),
\tilde{P}_D \to 
 \left(\begin{array}{ccc}1 & 0 &0 \\ 0 & -1& 0 \\ 0 & 0 &-1  \end{array}\right).
 \ee
Therefore, $D_N$ has only real representations.

$SU(2)$ is the universal covering group of $SO(3)$, and
has pseudo real and real irreps. 
$Q_{2N}$ is a finite subgroup of
$SU(2)$. It can be interpreted as
the covering group of $D_N$ in the sense
that the defining matrices
$\tilde{R}_{2N}$ and $\tilde{P}_Q$ for $Q_{2N}$ satisfy
\be
(\tilde{R}_{2N})^2=\tilde{R}_{N},~
(\tilde{P}_Q)^4=(\tilde{P}_D)^2={\bf 1},
\label{QandD}
\ee
where
\be
\tilde{R}_{2N}&= & 
\left(\begin{array}{ccc}\cos \frac{\theta_N}{2} & \sin 
\frac{\theta_N}{2}
\\ -\sin  \frac{\theta_N}{2} & \cos
 \frac{\theta_N}{2}  \end{array}\right),~
\tilde{P}_Q =
 \left(\begin{array}{ccc}i & 0 \\ 0 & -i
 \end{array}\right).
 \label{PQ}
 \ee
The set of $4N$ elements of $Q_{2N}$ is given by
 \be
{\cal G}_{Q_{2N}}=
\{\tilde{R}_{2N}, 
(\tilde{R}_{2N})^2,\dots,(\tilde{R}_{2N})^{2N}={\bf 1},
\tilde{R}_{2N} \tilde{P}_Q ,(\tilde{R}_{2N})^2 \tilde{P}_Q, 
\dots, (\tilde{R}_{2N})^{2N} \tilde{P}_Q=\tilde{P}_Q\}.
\ee

There exist only one- and two-dimensional irreps
for $D_N$ and $Q_{2N}$. For $Q_{2N}$,
there are $N-1$ different two-dimensional irreps,
which we denote by
\be
{\bf 2}_\ell,  ~~\ell=1,\dots,N-1.
\ee
${\bf 2}_\ell$ with odd $\ell$ is a pseudo real 
representation, while ${\bf 2}_\ell$ with even $\ell$ is a real 
representation, where ${\bf 2}_\ell$ with even $\ell$ is exactly ${\bf 2}_{\ell/2}$
of $D_N$. Under the  fundamental rotation
(i.e., a rotation  of $\theta_N$ which is defined in (\ref{thetaN})), 
${\bf 2}_\ell$ transforms with the matrix
\be
\tilde{R}_{2N}({\bf 2}_\ell) &=&(\tilde{R}_{2N})^\ell =
\left(\begin{array}{ccc}\cos (\ell\frac{\theta_N}{2}) & \sin 
(\ell\frac{\theta_N}{2})
\\ -\sin  (\ell \frac{\theta_N}{2}) & \cos
 (\ell\frac{\theta_N}{2}  ) \end{array}\right).
\ee
It is straightforward to calculate the Clebsch-Gordan 
coefficients for tensor products of irreps \cite{babu4}.
There exist four different one-dimensional irreps
of $Q_{2N}$.
Because of the relation (\ref{QandD}), each of them has
 a definite $Z_4$ charge.
Further, under the fundamental 
rotation, they either remain unchanged or change
their sign. Therefore, one-dimensional irreps 
can be characterized according to $Z_2\times Z_4$
charge:
\be
{\bf 1}_{+,0}, & &{\bf 1}_{-,0},~{\bf 1}_{+,2},~
{\bf 1}_{-,2} ~~\mbox{for}~~N=2,4,6,\dots,\\
{\bf 1}_{+,0}, & &{\bf 1}_{-,1},~{\bf 1}_{+,2},~
{\bf 1}_{-,3} ~~\mbox{for}~~N=3,5,7,\dots,
\ee
 where the ${\bf 1}_{+,0}$ is the true singlet of $Q_{2N}$,
 and only ${\bf 1}_{-,1}$ and ${\bf 1}_{-,3}$ 
 are complex irreps.
 Note that  all the real representations 
 of $Q_{2N}$ are exactly those of $D_N$,
 which is one of the reasons why we would like to call
 $Q_{2N}$ as the covering group of $D_N$.
 
\section{Field Theory with the dihedral invariance}
Let us now discuss how to construct
field theory models with a dihedral invariance.
We denote the five-dimensional coordinate by
\be
z^M &=&(x^\mu,y)~~\mbox{with}~~\mu=0,\dots,3.
\ee
The coordinates $y_i$  of the lattice sites
transform to $y_i'$ with $N\times N$ matrices of $D_N$,
which are given in (\ref{RN}) and (\ref{PD}).
Then it is natural to 
assume \footnote{$D_N$ may be understood as a twisted product of 
$Z_N$ and $Z_2$. Witten \cite{witten} has considered this $Z_N$
(the symmetry of the boundary of a deconstructed
disc) to solve the triplet-doublet splitting problem in GUTs.}
that the fields defined on
the lattice are irreps of $Q_{2N}$ which is the covering group of $D_N$.
That is \footnote{Nonabelian discrete family symmetries appearing in extra dimension
models of \cite{seidl1,altarelli}, for instance, are not directly related
to a symmetry of the extra dimension.},
\be
\phi(x,y)&\to & \phi'(x,y)= 
\tilde{Q}_{2N}~\phi(x, \tilde{D}_N^{-1}y),~
\tilde{Q}_{2N} \in Q_{2N} ~\mbox{and}~\tilde{D}_N \in D_N.
\label{trans1}
\ee
In Table 1 explicit expressions of the matrices
corresponding to the fundamental rotation and the parity
transformation are given, where we assume that the gauge fields
belong to the true singlet ${\bf 1}_{+,0}$.

\vspace{0.5cm}
\begin{center}
\begin{tabular}{|c|c|c|c|c|c|c|c|c|}
\hline
Irreps & ${\bf 1}_{+,0}$  &${\bf 1}_{+,2}$ 
&${\bf 1}_{-,0}$ & ${\bf 1}_{-,1}$  &${\bf 1}_{-,2}$ 
&${\bf 1}_{-,3}$ & ${\bf 2}_{2\ell -1}$ & ${\bf 2}_{2\ell }$
\\ \hline
rotation &$1$ & $1$ &$-1$ &
$-1$  & $-1$ &
$-1$ &$(\tilde{R}_{2N})^{2\ell-1}$
 &$(\tilde{R}_{2N})^{2\ell}$
\\ \hline
parity& $1$ 
& $-1$ & $1$ & $i$  & $-1$  
& $-i$  & $\tilde{P}_Q$ &$\tilde{P}_D$
\\ \hline
 reality & r & r & r & c& r & c & pr & r \\ \hline
\end{tabular}
\end{center}
\footnotesize
{\bf Table 1}. Explicit expressions of the matrices
corresponding to the fundamental rotation
(i.e., a rotation of $\theta_N$ given in (\ref{thetaN})) and the parity
transformation. $\tilde{R}_{2N}$, $\tilde{P}_Q$ and $\tilde{P}_D$
are given in (\ref{PQ}) and (\ref{PD2}), respectively, where
$\ell \in {\bf N}$ and $\le (N-1)/2$, r=real, c=complex, pr=pseudo real.
All the real irreps of $Q_{2N}$ are those of $D_N$.
Complex one-dimensional irreps exist only for $N=3,5,7,\dots$,
while the real one-dimensional irreps ${\bf 1}_{-,0}$ and ${\bf 1}_{-,2}$ 
exist only for $N=2,4,6,\dots$.
\normalsize

Given the details of the $Q_{2N}$ irreps, it is then straightforward 
to construct an invariant action \cite{hamed1,hill,hamed2}.
Supersymmetrization can also be straightforwardly done \cite{hamed2}.

\section{Orbifold boundary conditions and $Q_{2N}$ flavor symmetry}
In the case of a continuos extra dimension,
orbifold boundary conditions are used to suppress
unnecessary light fields and also to obtain four-dimensional chiral fields.
We shall discuss next how an internal 
$Q_{2N}$ flavor symmetry can appear even 
if orbifold boundary conditions break the dihedral invariance
(\ref{trans1}).
Let $\phi(x,y)$ be a generic field which satisfies the periodic
boundary condition,
$\phi(x,y)=\phi(x,y+N a)$.
Then the field $\phi(x,y)$ can be decomposed into the cosine and
sine modes:
\be
\phi(x,y) &=&
\frac{\phi(x)}{\sqrt{N}}+
\sum_{i=1}^{i_{\rm max}}\phi_{+,i}(x)~cos (k_i y)
+\sum_{i=1}^{i'_{\rm max}}\phi_{-,i}~(x)sin (k_i y)~,
\ee
where
\be
\phi(x) &=&\frac{1}{\sqrt{N}}\sum_{n=0}^{N-1} \phi(x,y_n),\\
k_i &=&  \frac{2\pi i}{a N}, i \in {\bf N},~
i_{\rm max}=\left\{\begin{array}{c}
i'_{\rm max}+1=N/2-1\\
i'_{\rm max}=(N-1)/2
\end{array}\right.~\mbox{for}~~
\left\{\begin{array}{c}
\mbox{even}~N\\
\mbox{odd}~N
\end{array}\right. .
\ee
$\phi(x)$ is the zero mode.
 As in the continuos case, we can drop the cosine or sine modes
 by imposing an appropriate boundary condition:
 Under the parity transformation (\ref{PD}), i.e.,
 \be
 y_0 & \to y'_0=y_0,~
 y_1 \to y'_1=y_{N-1}, ~
 \dots,~y_i \to y'_i=y_{N-i},\dots,
 \label{parity}
 \ee
 the zero mode  $\phi(x)$ and the cosines modes
are even, while the sine modes are odd.
 
Since the $D_N$ transformation mixes the cosine and sine modes,
the orbifold boundary conditions break the
dihedral invariance  explicitly. 
However, the $Q_{2N}$ invariant
construction of an action discussed in the previous section
ensures that
the $Q_{2N}$ invariance remains intact as a global,
internal symmetry. This is because 
there is no derivative with respect to $y$ is used
in the construction. So, the theory
with  orbifold boundary conditions
is invariant under the internal transformation
\be
\phi(x,y)&\to & \phi'(x,y)= 
\tilde{Q}_{2N}~\phi(x, y),~
\tilde{Q}_{2N} \in Q_{2N},
\label{trans2}
\ee
which should be compared with (\ref{trans1}).
The internal symmetry is nothing but a global flavor symmetry
based on $Q_{2N}$.

\section{An example}
In what follows, we would like to discuss a concrete model.
One of the successful Ans\" atze 
 for the quark mass matrices is
of a nearest neighbor interaction (NNI) type
\cite{weinberg,wilczek,fritzsch1}
 \be
 M &=&
  \left( \begin{array}{ccc}
0  & C &0 \\
\pm C & 0 &B\\
0 &  B'  &A\\
\end{array}\right).
\label{nni}
 \ee
In \cite{babu4}  it has been proposed to derive the mass matrix (\ref{nni}) 
solely from a dihedral symmetry, and concluded that
two conditions  should be met:
(i) There should be real as well as pseudo real 
nonsinglet representations,
and  (ii)  there should be  the up and down type
 Higgs $SU(2)_L$ doublets (type II Higgs).
The smallest finite group that allows both real  and pseudo real 
nonsinglet representations is $Q_6$ as we have
seen. Further, the Higgs sector
of the MSSM  fits  the desired Higgs structure.
Therefore, we assume supersymmetry in four dimensions.
The $D_3(S_3)$ model of \cite{kubo1} with a $Z_2$ symmetry
in the leptonic sector  is one of the most predictive models
for the leptonic sector. However, the $Z_2$ symmetry
in the quark sector is broken, so that the $Z_2$ symmetry
should be seen as an approximate symmetry in that model.
It was found, however, that this leptonic sector can
be reproduced in a supersymmetric $Q_6$ model  without
introducing an additional discrete symmetry 
into the leptonic sector \cite{babu4}.
In Table 2 we write the $Q_{6}$ assignment of the quark and
lepton chiral supermultiplets
\footnote{The same model exists for $Q_{2N}$
if $N$ is odd and a multiple  of $3$.}:
\vspace{0.5cm}
\begin{center}
\begin{tabular}{|c|c|c|c|c|c|c|}
\hline
 & $Q$ 
& $U^c,D^c,L,E^c,N^cH^u,H^d$ & $Q_3$  
& $U^c_3,D^c_3,H^u_3,H^d_3$ 
 & $L_3,E_3^c$  &  $N_3^c$
\\ \hline
$Q_6$ &${\bf 2}_1$ & ${\bf 2}_{2}$ &
${\bf 1}_{+,2}$  & ${\bf 1}_{-,1}$ & 
${\bf 1}_{+,0}$ & ${\bf 1}_{-,3}$  \\ \hline
\end{tabular}
\vspace*{5mm}
\end{center}

\footnotesize
{\bf Table 2}. $Q_{6}$ assignment 
of the matter supermultiplets.
$Q, Q_3, L, L_3$ and  $ H^u, H_3^u, H^d, H_3^d$
stand for $SU(2)_L$ doublets
supermultiplets for quarks, leptons and Higgs bosons, respectively.
Similarly, $SU(2)_L$ singlet
supermultiplets for quarks, charged leptons and neutrinos are denoted by
$U^c, U^c_3,D^c, D^c_3, E^c, E^c_3$ and $N^c, N^c_3$.
This is an alternative assignment to the one given in the footnote
of \cite{babu4}. The present assignment can more suppress
the proton decay \cite{itou1}.
The assignment for the mirror supermultiplets
can be simply read off from Table 2.
\normalsize

\noindent
We impose the following orbifold boundary conditions:
All the mirror chiral supermultiplets are odd
under  the parity transformation (\ref{parity}).
Similarly, the $N=1$ chiral supermultiplets,
which are the $N=2$ superpartners of
the  $SU(3)_C \times SU(2)_L \times U(1)_Y$ gauge supermultiplets, are also odd.
It is then clear that the zero modes of the
gauge, matter and Higgs supermultiplets
coincide with those of the supersymmetric $Q_6$ model of \cite{babu4},
and hence it is the  low energy effective theory.
The low energy Yukawa superpotential $W_Y$ is given by
\be
W_Y &=& W_Q+W_L,
\ee
where
 \footnote{ The Higgs sector of the model of \cite{babu4}
possesses a permutation symmetry $H^{u(d)}_{1}
 \leftrightarrow H^{u(d)}_{2}$, which ensures the stability of 
the  VEV $<H^{u(d)}_{1}>=
<H^{u(d)}_{2}>$. The resulting mass quark matrices
are  equivalent to (\ref{nni}).
 The leptonic sector given in \cite{kubo1} can be obtained
by the interchange $1 \leftrightarrow2$.} 
\be
W_Q &=&
Y_a^u Q_3 U_{3}^c H^u_3 + Y_b^u Q^T \sigma_1 U_{3}^c  H^u
 - Y^u_{b'} Q_3  U^{cT} i \sigma_2  H^u
+ Y^u_{c} Q^T  \sigma_1 U^c H^u_3\nn\\
& &+Y_a^d Q_3 D_{3}^c H^d_3 + Y_b^d Q^T \sigma_1 D_{3}^c  H^d
 - Y^d_{b'} Q_3  D^{cT} i \sigma_2  H^d 
+ Y^d_{c} Q^T  \sigma_1 D^c H^d_3,
\label{spotQ}\\
W_L &=&
Y_c^e f^{IJK}L_I E_J^cH_K^d
+Y_{b'}^e L_3 (H_1^d  E_1^c+  H^d_2 E_2^c)
+Y_{b}^e (L_1  H^d_1+L_2  H^d_2)E_3^c
\nn\\
& &+Y_{a}^\nu L_3 N_3^c  H^u_3
+Y_c^\nu f^{IJK}L_I N_J^cH_K^u
+Y_{b'}^\nu L_3 (H_1^u  N_1^c+  H^u_2 N_2^c),
\label{spotL}
\ee
and $f^{122}=f^{212}=f^{222}=-f^{111}=1$.
In \cite{babu4} it has been found
that by introducing a certain set of
gauge singlet Higgs supermultiplets
 it is possible to construct a Higgs sector in such a way that
CP phases can be spontaneously   induced.
Therefore, all the parameters appearing in the 
Lagrangian including the soft
supersymmetry breaking (SSB) sector are real.
Consequently, no $CP$ violating processes
induced by SSB terms are possible in this model,
satisfying the most stringent experimental constraint
coming from the EDM of the neutron and the electron \cite{edm}.
Since the Higgs sector is also $Q_{6}$ invariant, it is straightforward
to derive it from dimensional deconstruction.
Consequently, 
the quark  sector contains only 8 real parameters with 
one independent phase to describe  the quark masses  and  their mixing,
and the leptonic  sector contains only 6 real parameters
with one independent phase 
to  describe  12 independent physical parameters.
Predictions  in the $|V_{ub}|-\sin 2 \phi_1$ planes are
shown in Fig.~2, while Fig.~3 shows the predictions 
in the $\sin 2 \phi_1-\phi_3$ planes.
%\begin{center}
\begin{figure}[htb]
 \parbox{6cm}{%
\includegraphics*[width=0.4\textwidth]{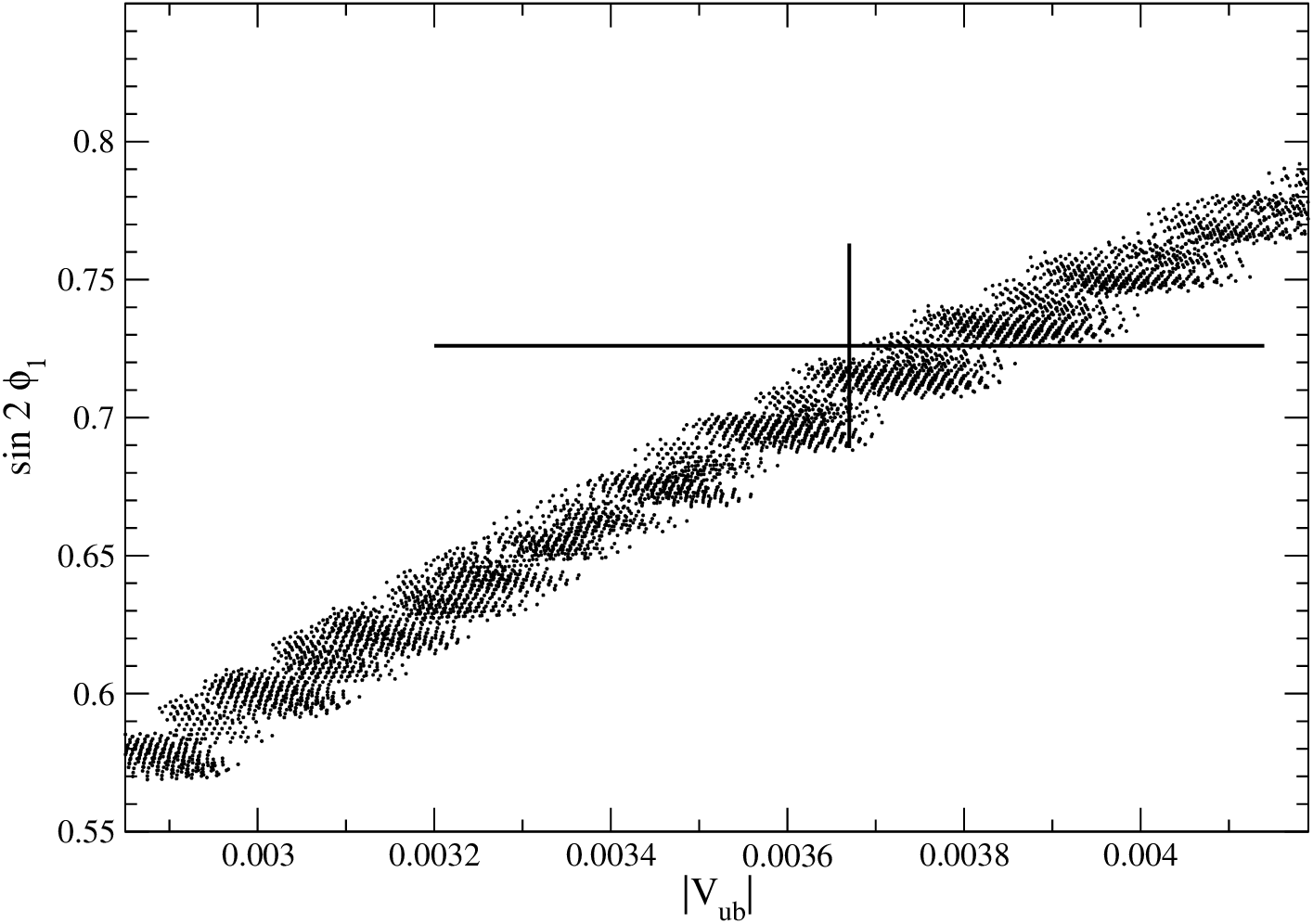}
\caption{\label{fig2}\footnotesize
Predictions in the $|V_{ub}|-\sin2\phi_1$ plane.
The uncertainties result from those in the quark masses
and in  $|V_{us}|$ and $|V_{cb}|$, where
we have used
$|V_{us}|=0.2240\pm 0.0036$ and 
$|V_{cb}|=(41.5\pm 0.8)\times 10^{-3}$ \cite{kim1}.
The vertical and horizontal lines correspond to
the experimental values,
$\sin 2 \beta(\phi_1)=0.726\pm 0.037$ and  
$|V_{ub}|=(36.7\pm 4.7)\times 10^{-4}$ \cite{pdg,hfag}.
\normalsize }}
\hspace{1cm}
 \parbox{6cm}{%
\includegraphics*[width=0.4\textwidth]{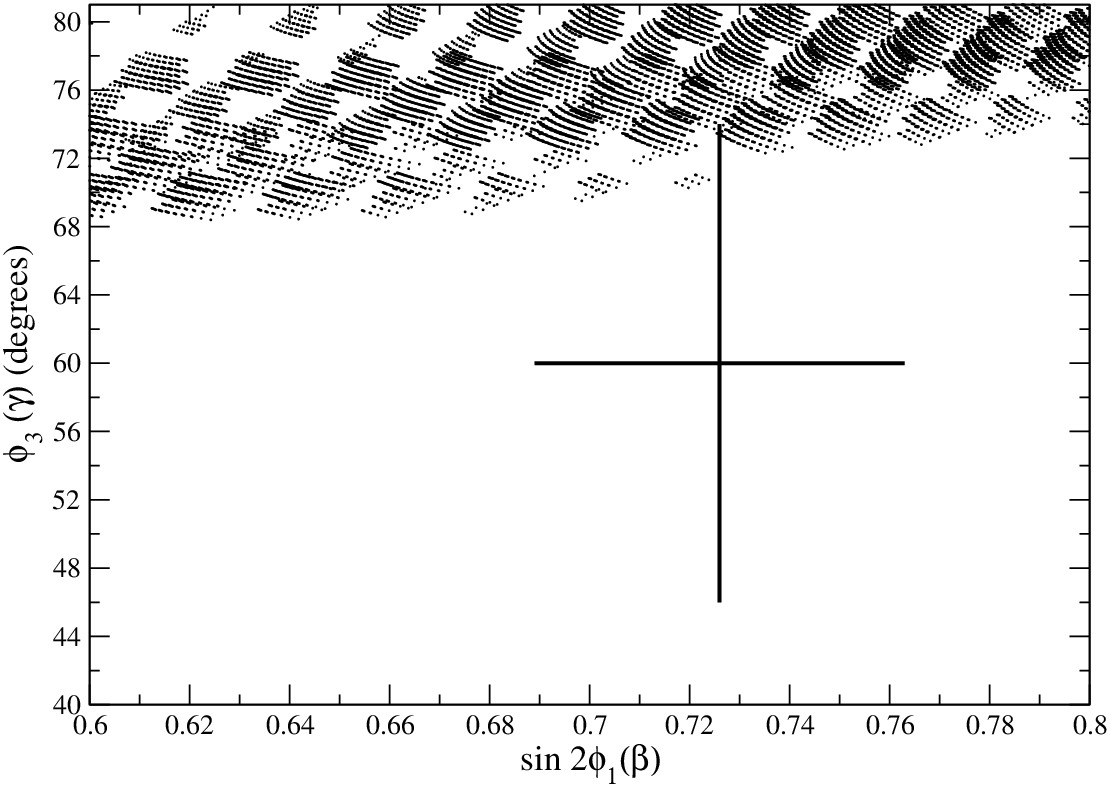}
\caption{\label{fig3}\footnotesize
Predictions in the $\sin2\phi_1-\phi_3$ plane.
The vertical and horizontal lines correspond to
the experimental values,
$\sin 2 \phi_1(\beta)=0.726\pm 0.037$ and  
$\phi_3=(60^o\pm 14^o) $ \cite{pdg,hfag}.
\normalsize}}
\end{figure}
%\end{center}
As we can see from Fig.~2 and 3, with accurate 
 measurements of
the Cabibbo-Kobayashi-Maskawa matrix elements, the predictions
could be tested.

The predictions in the leptonic sector are
summarized as follows \footnote{Large mixing of neutrinos may be
obtained in dimensional deconstruction models in a
different mechanism. See, for instance, \cite{seidl1,balaji,enkhbat}.}:
\begin{enumerate}
\item
Inverted neutrino mass spectrum, i.e.,
$m_{\nu_3}  <  m_{\nu_1}, m_{\nu_2}$.

\item
$m_{\nu_2}^2/\Delta m_{23}^2 =
\frac{(1+2 t_{12}^2+t_{12}^4-r t_{12}^4)^2}
{4  t_{12}^2 (1+t_{12}^2)(1+t_{12}^2-r t_{12}^2)\cos^2 \phi_\nu}
-\tan^2 \phi_\nu~~(r=\Delta m_{21}^2/\Delta m_{23}^2,  t_{12}=\tan\theta_{12})$,
where $\phi_\nu$ is an independent phase.

\item
$\sin\theta_{13}  \simeq m_e/\sqrt{2} 
m_\mu\simeq 3.4 \times 10^{-3}$ and
$\tan\theta_{23}  
\simeq 1-(m_e/\sqrt{2}m_\mu)^2
=1-O(10^{-5})$.

\item
The prediction of $<m_{ee}>$  is shown in Fig.~4.
\end{enumerate}
\begin{center}
\begin{figure}[htb]
\includegraphics*[width=0.4\textwidth]{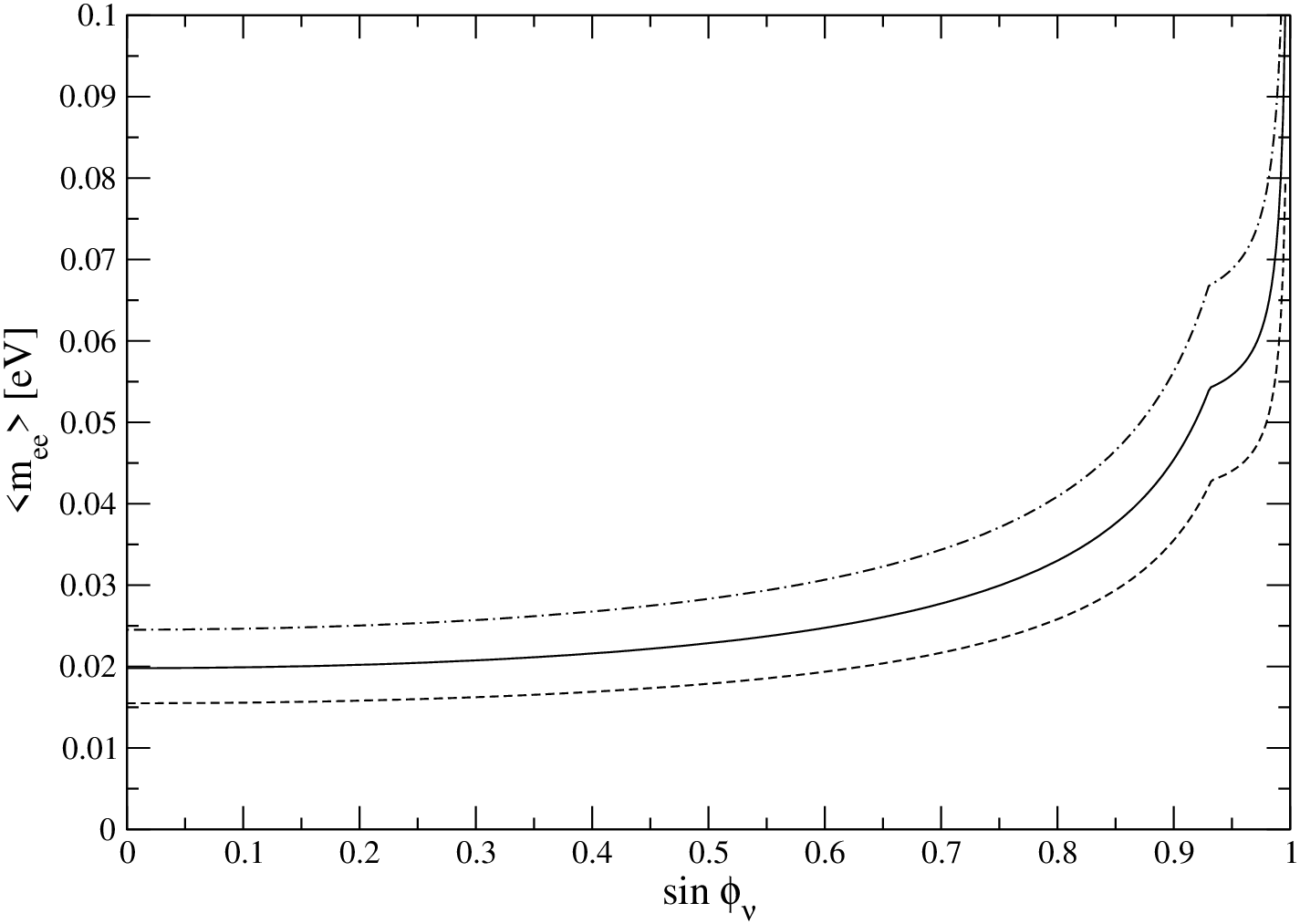}
\caption{\label{fig4}
\footnotesize
The effective Majorana mass $<m_{ee}>$ as a function of
$\sin \phi_\nu$ with 
$\sin^2\theta_{12}=0.3$ and 
 $\Delta m_{21}^2=6.9 \times10^{-5}$ eV$^2$ \cite{maltoni3}.
 The dashed, solid and dot-dashed lines stand for 
$\Delta m_{23}^2=1.4, 2.3$ and $ 3.0 \times 10^{-3}$ eV$^2$,
respectively.}
\end{figure}
\end{center}
\normalsize
We emphasize that 
the smallness of $\sin\theta_{13}$ and the almost maximal
mixing of the atmospheric neutrinos are consequences
of the $Q_6$ flavor symmetry. $\sin\theta_{13}$
in the present model may be too small
to be measured in a laboratory experiment
\cite{minakata}, but the tiny deviation from zero
($\sin^2\theta_{13} \simeq m_e^2/2 m_\mu^2\simeq 10^{-5}$)
are important in supernova neutrino oscillations\cite{ando}.

\section{Conclusion}
In this letter we have looked for a possible origin of
dihedral symmetries.
It has been recently
realized that a flavor symmetry based on a 
dihedral group can be used to 
soften the flavor problem of the SM and the MSSM.
We have considered
an extra dimension compactified on a closed chain,
which is assumed to have the form of a regular polygon.
Since the symmetry group of the regular polygon is
the dihedral group $D_N$,
we assumed that the fields are irreps of the  covering group
of $D_N$, which is 
the binary dihedral group $Q_{2N}$.
The construction of an action with the dihedral  invariance
is straightforward, and moreover we found that
the $Q_{2N}$ symmetry remains as an intact, internal flavor
symmetry even if the original dihedral invariance is
broken by orbifold boundary conditions.
We hope that with our finding 
we can come closer to a deep understanding of
 the origin of a flavor symmetry based on
a nonabelian finite group.


\begin{thebibliography}{99}
\bibitem{susy}    
S.P.~ Martin, hep-ph/9709356.

\bi{frampton1}
 P.H.~Frampton and T.W.~Kephart,
Int. J. Mod. Phys. {\bf A10,} 4689 (1995);
P.H.~Frampton and T.W.~Kephart,
Phys. Rev. {\bf D64,} 086007 (2001). 

\bi{ma1}
E.~Ma and G.~Rajasekaran, Phys. Rev. {\bf D64,} 113012 (2001);
E.~Ma, Mod. Phys. Lett. {\bf A17,} 627 (2002); 2361 (2002) .

\bi{babu1}
K.S.~Babu, E.~Ma and J.W.F.~Valle,
Phys. Lett. {\bf B552,} 207 (2003).

\bi{seidl1}
G.~Seidl, hep-ph/0301044.

\bi{kubo1}
J.~Kubo, A.~Mondrag\'on, M.~Mondrag\'on and
E.~Rodr\' iguez-J\' auregui,  Prog. Theor. Phys.
{\bf 109,} 795 (2003);
J.~Kubo, Phys. Lett. {\bf B578,} 156 (2004).

\bi{grimus2}
W.~Grimus and L.~Lavoura,  Phys. Lett. {\bf B572,}  76 (2003).

\bi{kobayashi1}
T.~Kobayashi, S.~Raby and R-J.~Zhang, Phys. Lett. {\bf B593,}
262 (2004);
Nucl. Phys. {\bf  B704,}3 (2005).

\bi{ma5}
S-L. Chen, M. Frigerio and E. Ma,
Phys. Rev. {\bf D70,} 073008 (2004).

\bi{grimus3}
W.~Grimus, A.S.~Joshipura, S.~Kaneko, L.~Lavoura and M.~Tanimoto,
JHEP {\bf 0407,} 078 (2004).

\bi{frigerio}
M.~Frigerio, S.~Kaneko, E.~Ma and  M.~Tanimoto, 
Phys. Rev. {\bf D71,} 011901 (2005).



\bi{babu4}
K.S.~Babu and J.~Kubo, Phys. Rev. {\bf D71,} 056006 (2005).

\bi{ma6}
 L.~Lavoura and E.~Ma, hep-ph/0502181;
S-L.Chen and E.~Ma, hep-ph/0505064.

\bi{hagedorn}
C.~Hagedorn and W.~Rodejohann, hep-ph/0503143. 

\bi{caravaglios}
F.~Caravaglios and  S.~Morisi, hep-ph/0503234.

\bi{grimus4}
W.~Grimus and L.~Lavoura, hep-ph/0504153.

%%%%%
\bi{kobayashi2}
T.~Kobayashi, J.~Kubo and H.~Terao,
Phys. Lett. {\bf B568,} 83 (2003).

\bi{choi}
Ki-Y.~Choi, Y.~Kajiyama,
J.~Kubo and H.M.~Lee,  Phys. Rev. {\bf D70,} 055004 (2004).
%%%%%

\bi{hall2}
L.J.~Hall and H.~Murayama,
Phys. Rev. Lett. {\bf 75,} 3985 (1995)  ;
C.D.~Carone, L.J.~Hall and H.~Murayama,
Phys. Rev. {\bf D53,} 6282 (1996).


\bi{hamaguchi}
K.~Hamaguchi, M.~Kakizaki and M.~Yamaguchi, 
Phys. Rev. {\bf D68,} 056007 (2003).

\bi{babu2} 
K.S.~Babu, T.~Kobayashi and J.~Kubo, 
Phys. Rev. {\bf D67,} 075018 (2003).


\bi{maekawa}
N.~Maekawa and T.~Yamashita,
JHEP {\bf  0407,} 009 (2004); T.~Yamashita, hep-ph/0503265.

\bi{king2}
 G. G. Ross, L.~Velasco-Sevilla
 and  O.~Vives, Nucl. Phys. {\bf B692,} 50 (2004).

%%%%%%%
\bi{hamed1}
N.~Arkani-Hamed, A.G.~Cohen and H.~Georgi,
Phys. Rev. Lett. {\bf 86,} 4757 (2001);
Phys. Lett. {\bf B513,} 232 (2001).

\bi{hill}
C.T. Hill, S.~Pokorski and J.~Wang,
 Phys. Rev. {\bf D64,} 105005 (2001).
 
 \bi{hamed2}
 N.~Arkani-Hamed, A.G.~Cohen and H.~Georgi,
JHEP {\bf  0207,} 020  (2002).

\bi{witten}
E.~Witten, hep-ph/0201018.

%%%%%%%

\bi{altarelli}
 G.~Altarelli and F.~Feruglio, hep-ph/0504165.

%%%%%
\bi{weinberg}S.~Weinberg,
in {\em Transactions of the New York Academy of Sciences}, 
Series II, Vol. 38, 185 (1977).

\bi{wilczek} F.~Wilczek and A.~Zee,
Phys. Rev. Lett. {\bf 42,} 421 (1979).

\bi{fritzsch1}H.~Fritzsch,
Phys. Lett. {\bf B 73,} 317 (1978);  Nucl. Phys. {\bf B155,} 189 (1979). 

\bi{itou1}
E.~Itou, Y.~Kajiyama and J.~Kubo, in preparation.

\bi{balaji}
K.R.S.~Balaji, M.~Lindner and G.~Seidl,
Phys. Rev. Lett. {\bf 91,} 161803 (2003).

\bi{enkhbat}
 T.~Enkhbat and G.~Seidl, hep-ph/0504104;
T.~Hallgren, T.~Ohlsson and G.~ Seidl,
JHEP {\bf  0502,} 049 (2005).

\bibitem{edm} 
F.~Gabbiani, E.~Gabrielli, A.~Masiero and L.~Silvestrini,
Nucl.~Phys. {\bf B477,}  321 (1996);
S.~Abel, S.~Khalil and O.~Lebedev,
Nucl.~Phys. {\bf B606,}  151 (2001); M.~Endo, M.~Kakizaki and 
M.~Yamaguchi, Phys.~Lett.~{\bf 583,} 186 (2004); 
J.~Hisano and  Y.~Shimizu,  Phys. Rev. {\bf D70,} 093001(2004).


\bi{kim1}
H.D.~Kim, S.~Raby and L.~Schradin, hep-ph/0401169.

\bi{pdg}
Particle Data Group,
Phys. Lett . {\bf B592,}  1 (2004).

\bi{hfag} Heavy Flavor Averaging Group (HFAG), hep-ex/0412073.

\bi{maltoni3}
M.~Maltoni, T.~Schwetz, M.A.~T\`ortalo and J.W.F.~Valle,
New J. Phys. {\bf 6,} 122 (2004).

\bi{ando}S.~Ando and K.~Sato,
Phys. Rev. {\bf D68,} 023003 (2003);  JCAP {\bf 0310,} 001(2003).

\bi{minakata}
H.~Minakata, H.~Sugiyama, O.~Yasuda, K.~Inoue and  F.~Suekane,
Phys. Rev. {\bf D68,} 033017 (2003); Erratum-ibid. {\bf D70,} 059901 (2004);
O.~Yasuda, hep-ph/0309333;
H.~Minakata, Nucl. Phys. {\bf B (Proc. Suppl.) 137,} 74 (2004);
H.~Sugiyama, O.~Yasuda, F.~Suekane and 
G.A.~Horton-Smith, hep-ph/0409109.



\end{thebibliography}
\end{document}